\documentclass{sigchi}

\setcopyright{none}
\renewcommand{\copyrightnotice}{}

\usepackage{balance}       
\usepackage{graphics}      
\usepackage[T1]{fontenc}   
\usepackage{txfonts}
\usepackage{mathptmx}
\usepackage[pdflang={en-US},pdftex]{hyperref}
\usepackage{color}
\usepackage{booktabs}
\usepackage{textcomp}

\usepackage[%
PlaceholderOption]{drafthelper}

\newenvironment{ParticipantQuote}%
{\begin{small}\begin{list}{0}{%
\setlength{\topsep}{0.0pt}%
\setlength{\itemsep}{0.6pt}%
\setlength{\leftmargin}{0.3in}%
\setlength{\itemindent}{0.0in}%
\setlength{\rightmargin}{0.1in}%
\setlength{\labelsep}{0.25em}}}%
{\end{list}\end{small}}

\newenvironment{RQList}%
{\begin{list}{0}{\setlength{\leftmargin}{0.2in}\setlength{\itemindent}{0.0in}\setlength{\labelsep}{0.25em}\setlength{\itemsep}{0pt}}}%
{\end{list}}

{\begin{list}{$\bullet$}{\setlength{\leftmargin}{0.05in}\setlength{\labelsep}{0.5em}}}%
{\end{list}}

\newcommand{\Code}[1]{\texttt{#1}}

\usepackage{microtype}        
\usepackage{ccicons}          

\usepackage{todonotes}

\def\plaintitle{Wandercode: An Interaction Design for Code Recommenders to Reduce Information Overload, Ease Exploration, and Save Screen Space}

\def\emptyauthor{}
\def\plainkeywords{Programming; code navigation; recommender system}

\makeatletter
\def\url@leostyle{%
  \@ifundefined{selectfont}{
    \def\UrlFont{\sf}
  }{
    \def\UrlFont{\small\bf\ttfamily}
  }}
\makeatother
\def\pprw{8.5in}
\def\pprh{11in}

\definecolor{linkColor}{RGB}{6,125,233}
\hypersetup{%
  pdftitle={\plaintitle},
  pdfauthor={\emptyauthor},
  pdfkeywords={\plainkeywords},
  pdfdisplaydoctitle=true, 
  bookmarksnumbered,
  pdfstartview={FitH},
  colorlinks,
  citecolor=black,
  filecolor=black,
  linkcolor=black,
  urlcolor=linkColor,
  breaklinks=true,
  hypertexnames=false
}

\begin{document}

\title{\plaintitle}

\numberofauthors{3}
\author{%
  \alignauthor{Austin Z. Henley\\
    \affaddr{Carnegie Mellon University}\\
    \email{azhenley@cmu.edu}}\\
  \alignauthor{David Shepherd\\
    \affaddr{Louisiana State University}\\
    \email{dshepherd@lsu.edu}}\\
  \alignauthor{Scott D. Fleming\\
    \affaddr{University of Memphis}\\
    \email{Scott.Fleming@memphis.edu}}\\
}

\maketitle


\begin{abstract}
In this paper, we present Wandercode, a novel interaction design for recommender systems that recommend code locations to aid programmers in software development tasks. In particular, our design aims to improve upon prior designs by reducing information overload, by better supporting the exploration of recommendations, and by making more efficient use of screen space. During our design process, we developed a set of design dimensions to aid others in the design of code recommenders. To validate our design, we implemented a prototype of our design as an Atom code editor extension with support for the Java programming language, and conducted an empirical user evaluation comparing our graph-based Wandercode design to a control design representative of prior list-based interaction designs for code recommenders. The results showed that, compared with the control design, Wandercode helped participants complete tasks more quickly, reduced their cognitive load, and was viewed more favorably by participants.
\end{abstract}

%





\section{Introduction}
\label{sec-introduction}

\Intent{Providing recommendations to users is helpful.}
Contemporary software users are often faced with the problem of making decisions without sufficient information, and \emph{recommender systems} are being applied more and more to address such problems~\cite{Resnick1997CACM}.
Recommender systems aim to automatically predict the value of a piece of information to a particular user in a particular context and to present the user with the items of greatest predicted value (i.e., \emph{recommendations}).
The designs of recommender systems often vary depending on the application, and researchers have investigated designs for a wide variety of domains, including recommendations for learning software~\cite{Matejka2009UIST}, planning trips~\cite{Kim2009CHI}, finding online discussions~\cite{Chen2011CHI, Chen2010CHI}, blogging~\cite{Dugan2010CHI}, fixing software errors~\cite{Hartmann2010CHI}, creating mix tapes~\cite{Hansen2009CHI}, and finding friends on social media~\cite{Chen2009CHI}.
%

\Intent{A particularly interesting environment is that of software development, to recommend relevant locations in source code to the programmer.}
A particularly promising domain for recommender systems is that of software development, where programmers need to find the code relevant to their tasks---often from among many thousands of interconnected code modules.
The consequences of this challenge are all too real for modern programmers. 
For example, studies of programmers engaged in debugging found that programmers spent 30--50\% of their time navigating code~\cite{Ko2005ICSE, Piorkowski2013CHI}, and this excessive time and effort has been largely attributed to programmers accidentally wasting time investigating irrelevant code~\cite{Ko2006TSE, Piorkowski2016FSE, Piorkowski2017VLHCC}.
Effectively delivering programmers recommendations of task-relevant code could make a substantial impact, reducing time wasted on unhelpful navigation and enhancing programmer productivity overall.

To address this problem, software engineering researchers have investigated a number of recommender systems (e.g., Mylar~\cite{Kersten2006FSE} and Stacksplorer~\cite{Karrer2011UIST}); however, to date, little research has investigated effective \emph{interaction designs} for delivering recommendations to programmers---an aspect critical to the success of a recommender system approach~\cite{Murphy-Hill2014RSSE}.
%
%
%
In particular, prior research on recommender systems for programmers has focused on what algorithms and factors to use to make recommendations~\cite{Murphy-Hill2014RSSE, Robillard2010Software}.
However, recommendations will be useful to programmers only if they can effectively and efficiently understand what the recommendation refers to and why the recommendation was selected.
Moreover, programmers will have difficulty adopting new recommender tools unless the interactions for receiving and understanding recommendations integrate cleanly into the already complex and cognitively taxing activities of modern programmers.
%

Existing recommender systems exhibit several apparent interaction design problems that may hinder their effectiveness for programmers.
To date, the predominant interaction designs have been \emph{list based}, displaying lists of recommendations that are sorted by relevancy (e.g.,~\cite{Piorkowski2012CHI, Singer2005ICSM, Warr2007ICSE}).
Unfortunately, these list-based designs have tended (1) to exacerbate the problems of information overload already faced by modern programmers, (2) to make exploring recommendations difficult, and (3) to make inefficient use of screen space.
Seeking to improve upon list-based designs, researchers have more recently proposed graph-based interaction designs that better capture the interrelationships between recommendations (e.g.,~\cite{Augustine2015ICSE, LaToza2011VLHCC}); however, these designs have also tended to worsen the problem of information overload problem and to consume even more screen space than the list-based designs.

To address problems with the prior recommender systems, we introduce a novel graph-based design, \emph{Wandercode}.
In particular, our design seeks to reduce information overload, to better support exploring recommendations, and to utilize screen space more efficiently.
To reduce information overload, it provides a more focused set of recommendations, and reduces the supporting cues to those expected to be most useful.
To support exploration, it provides features to incrementally expand related subsets of recommendations and to fix the visualization in place to prevent disorientation from auto-updates.
To better utilize screen space, Wandercode overlays the graph of recommendations on top of the code editor.
This overlay both saves space and obviates the need for window management activities that using a window or a pane would require.
%

\Intent{To validate the benefits that Wandercode provides, we ran an empirical study.}
To validate the benefits that Wandercode provides, we conducted an empirical user evaluation comparing our Wandercode design with a control list-based design.
To conduct the study, we implemented Wandercode as an extension to the Atom code editor with support for the Java programming language.
For the control design, we similarly implemented a representative list-based design modeled after existing tools.
The study was within-subjects and consisted of 10 programmers performing tasks that required navigating and understanding code modules in a large open source code base.
In particular, the evaluation investigated three research questions:

\begin{RQList}

\item[RQ1:] Do programmers using Wandercode spend less time during tasks?

\item[RQ2:] Do programmers using Wandercode have lower cognitive load during tasks?

\item[RQ3:] Do programmers have favorable opinions of Wandercode?

\end{RQList}

\Intent{Our work makes several key contributions.}
Our work makes the following key contributions: 

\begin{itemize}
\setlength\itemsep{-0.1em}


\item A novel interaction design for recommending code to programmers, Wandercode, that is noteworthy as the first programming tool design to overlay a code editor with an interactive graph visualization.

\item A prototype implementation of Wandercode for Java as an Atom code editor extension.

\item A set of design dimensions for code recommenders identified during our design process.

\item The findings of an empirical evaluation of Wandercode that found that the design helped programmers complete tasks faster, lowered cognitive load, and was viewed more favorably by participants.

\end{itemize}

\DraftPageBreak{}


\section{Background \& Related Work}
\label{sec-relatedwork}

%

\subsection{Contemporary Programming Environments}

\Intent{Code is huge, but programmers have to understand a lot of it.}
Modern software may comprise millions of lines of code that are organized into hundreds of thousands of interrelated modules.
In an effort to comprehend how the code works, programmers spend considerable amount of time navigating the huge set of code.
For example, one study found that programmers spent 35\% of their time on the mechanics of navigating the code~\cite{Ko2005ICSE}.
In another study, programmers spent 50\% of their time foraging for information~\cite{Piorkowski2013CHI}.
Evidence suggests that a substantial amount of this time is spent investigating irrelevant code~\cite{Ko2006TSE}, which may be caused by difficulty in predicting the value of a navigation~\cite{Piorkowski2016FSE}.

\begin{figure*}
\centering
\includegraphics[scale=1.0]{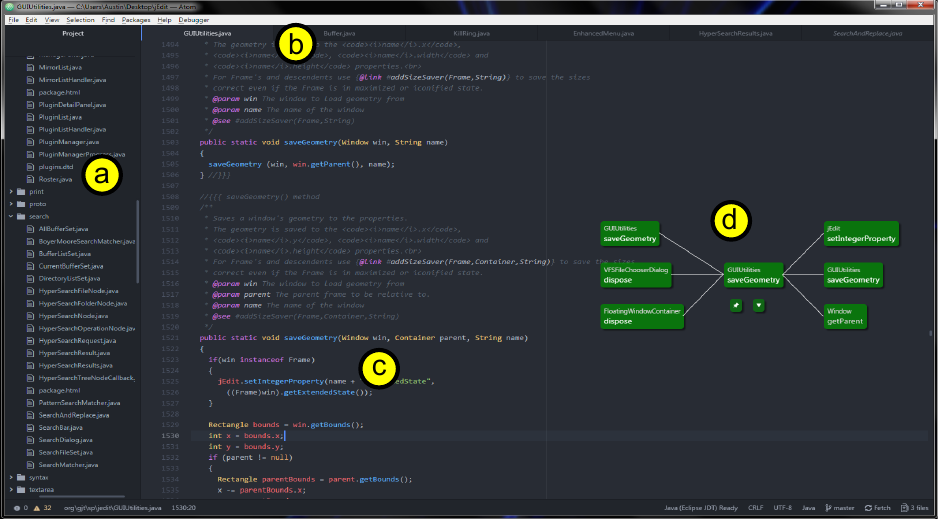}

\vspace{-1pt}
\caption{%
The Atom code editor with the (a) project explorer and (b) tabs of open documents. 
When the programmer moves the text cursor into (c) a method, then (d) Wandercode's graph of recommendations is displayed.
%
%
}
\label{fig-wandercode}
\end{figure*}

\Intent{To navigate code, programmers use these features of code editors.}
To navigate code, contemporary programmers often use programming environments, such as Eclipse, Visual Studio, and Atom, which provide affordances for navigating and searching code.
The Atom code editor, a particularly popular code editor with over one million active users~\cite{Dohm2016Blog}, is depicted in \CalloutFigure{fig-wandercode}.
Using the project explorer (\CalloutFigure{fig-wandercode}a) the programmer can see every code file in the project, and when clicked, the file opens in a tab document (\CalloutFigure{fig-wandercode}b) that displays the code (\CalloutFigure{fig-wandercode}c).
Additionally, code editors often provide many different views that can display a variety of information to the programmer (e.g., Eclipse's Outline View or Atom's Minimap feature).
%
%
Although researchers have proposed more radical user interfaces for code editors to enable more efficient code navigation (e.g., Code Bubbles~\cite{Bragdon2010CHI} and Patchworks~\cite{Henley2014CHI}), these designs have not seen substantial adoption.

\subsection{Recommender Systems for Code Locations}

\Intent{Recommender systems for code locations...}
Since software projects are so large, recommender systems for code can provide programmers suggestions of code modules that may be of interest to the programmer's current task.
%
%
In this context, a recommendation is any information that is ``estimated to be valuable for a software engineering task''~\cite{Robillard2010Software}.
%
Such recommender systems have utilized a variety of factors, often multiple, to produce recommendations of code.
Notable factors include code structure~\cite{Augustine2015ICSE, Hill2007ASE, Holmes2005ICSE, Karrer2011UIST, Kramer2012CHI, LaToza2011VLHCC, Warr2007ICSE}, natural language~\cite{Hill2007ASE, Ye2002ICSE}, navigation history~\cite{Kersten2006FSE, Piorkowski2012CHI, Singer2005ICSM}, code edits~\cite{Kersten2006FSE}, collaborative information~\cite{DeLine2005VLHCC, Singer2005ICSM}, and project documents~\cite{Cubranic2003ICSE}.
In this work, we will focus on using code structure as the primary factor in recommending code.
%

\subsection{Existing Interaction Designs and Their Drawbacks}

\Intent{The predominant way of displaying recommendations is in the form of a list, but this approach has issues.}
The predominant interaction design in prior work for providing recommendations to programmers has been a mixed-initiative list of code locations (e.g., Navtracks~\cite{Singer2005ICSM}, Suade~\cite{Warr2007ICSE}, and the PFIS recommender~\cite{Piorkowski2012CHI}).
Such \emph{list-based} designs typically display the recommendations in a sorted list (from greatest to least predicted relevance), denoting each code location with a name (often the name or signature of the code function/module/component).
The design is mixed initiative in that the list automatically updates as the programmer opens and interacts with different code locations in their code editor.
If the programmer takes interest in a recommendation, they may take the initiative, interacting with the list and exploring it further.
Each list item typically hyperlinks to a location in the code such that clicking the link will cause the code location to open in the code editor.
Additional cues may be included with each item to help clarify what the item is and why it was chosen (e.g., a list of relevant keywords~\cite{Piorkowski2012CHI}).
%

However, these list-based interaction designs have several apparent drawbacks.
First, several aspects of list-based designs contribute to \emph{information overload}.
Recommendation lists have tended to be long (e.g., 20 or more items being common) such that it would take substantial time to review all items.
Moreover, such long lists exceed working memory limits~\cite{Cowan2001, Miller1956}, and thus, may increase cognitive load in processing the list.
To make matters worse, it is often difficult for programmers to tell what each item is and why it was chosen.
Attempts to address this by adding cues results in even more information to process, and the relative effectiveness of different types of cues has received little or no study.

Second, \emph{exploration of recommendations} has not been well supported in list-based designs.
It is often difficult to tell the relationships between recommended code locations and  to explore the code locations related to a particular recommended location of interest.
Understanding the inter-dependencies of software components is key to effective program understanding~\cite{Fritz2014FSE}; however, such information is not conveyed in a list visualization.
Exploration has been further made difficult by list-based designs tendencies to repopulate the list in response to programmer exploration interactions.
Such updates may be jarring to a programmer, straining their working memory and depriving them of the ability to leverage spacial memory, since the recommendations update automatically.

Third, \emph{screen space} in modern programming environments is already crowded with numerous views (a code editor, a project explorer, a console, etc.), and list-based designs add yet another view.
Adding this additional list view hinders the usability of other views by compacting them or outright removing them.
Moreover, the increased crowding can lead the programmer to expend additional effort on tedious window management activities trying to organize their development environment.

In an attempt to address the issues with list-based designs, researchers have begun to investigate \emph{graph-based} interaction designs (e.g., Prodet~\cite{Augustine2015ICSE} and Reacher~\cite{LaToza2011VLHCC}).
Such designs typically present recommendations within a box-and-line graph visualization.
Prior implementations have used the boxes to hold the recommended code locations and used the lines to represent \emph{call dependencies} between the code locations.
Call dependencies are a key code-structural relationship that programmers seek to understand during coding tasks.
In most other respects, the graph-based design is similar to the list-based design: the recommendations are automatically updated in a mixed-initiative fashion, the recommendations hyperlink to code locations, and additional cues are often added to clarify what the recommendation is and why it was recommended.
%
%
%
%

Two notable recommender systems that display a graph of recommendations are Prodet~\cite{Augustine2015ICSE} and Reacher~\cite{LaToza2011VLHCC}.
Prodet provides a call graph visualization, where each recommendation is a code method that is related to the code method that the programmer is currently investigating in the code editor~\cite{Augustine2015ICSE}.
The graph is displayed in a pane at the bottom of the code editor and filters the recommendations based on heuristics such as textual similarity, code structure, and navigation history.
In an empirical study lasting six weeks, Prodet was found to double the number of structural navigations that programmers made, which are considered to be a highly efficient means of navigation~\cite{Augustine2015ICSE}.
Another tool, Reacher, displays a similar graph at the bottom of the code editor~\cite{LaToza2011VLHCC}.
Reacher also provides features to do structural searches on the graph, and it displays additional relationship information on the graph (e.g., this method is called inside of a loop or in a conditional statement).
An initial lab study of Reacher found that participants completed tasks 5.6 times more successfully and did so in 35\% less time than participants without Reacher~\cite{LaToza2011VLHCC}.

Although the graph-based designs have made headway in addressing some of the issues with list-based designs, they have left other issues unaddressed or even exacerbated.
A key improvement is that the graph-based design makes interrelationships between recommended code locations explicit.
This explicitness aids exploration by elegantly conveying code dependencies, and may have been the key factor in improving programmer performance in the prior studies.
However, jarring automatic updating of the recommendations and the graph remain an exploration barrier in the prior graph-based designs.
Furthermore, the information overload problem seems to have gotten worse, with graph-based designs including not only the same sorts of cues that list-based designs include, but adding even more visual noise in the form of colors and icons.
Likewise, the graph-based designs have worsened the crowding of screen space, because the graph visualization tends to require more space than does the comparatively compact list-based designs.

\DraftPageBreak{}


\section{Wandercode}
\label{sec-wandercode}

\begin{figure}
\centering
\includegraphics[scale=1.0]{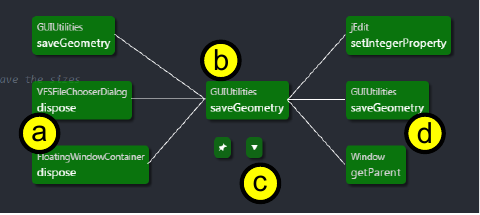}

\vspace{-1pt}
\caption{%
Close-up of Wandercode's graph of recommendations.
The (a) three recommendations on the left \emph{call} the (b) currently selected method of code.
On the right (d) shows three recommendations that get \emph{called} by the currently selected method.
(c) Two buttons: one to \emph{pin} the graph, preventing it from updating, and the other to \emph{expand} the graph, showing more recommendations.
Each recommendation shows the method name and class name, and clicking it will navigate to that code location.
}
\label{fig-wandercode-graph}
\end{figure}

\begin{figure}
\centering
\includegraphics[scale=1.0]{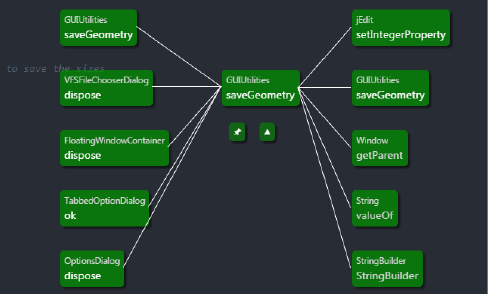}

\vspace{-1pt}
\caption{%
Close-up of Wandercode's graph of recommendations that has been \emph{expanded}.
Wandercode will display up to 10 recommendations in this mode (i.e., 5 callers and 5 callees)
This feature is enabled by clicking the expand button, shown in \CalloutFigure{fig-wandercode-graph}c, and clicking it again will collapse the graph back.
}
\label{fig-wandercode-expanded}
\end{figure}

\Intent{Wandercode aims to overcome issues of other recommenders and we made these design choices.}
To overcome apparent issues with both traditional list-based and graph-based recommender systems for code, we propose the Wandercode tool concept.
It does this by utilizing a graph visualization of recommendations using a mixed-initiative interaction design.
We implemented Wandercode as an extension to the Atom code editor~\cite{Atom}, shown in \CalloutFigure{fig-wandercode}.
During our design process, we identified three primary design goals for Wandercode to overcome limitations of other recommender systems: (1) utilize screen space more efficiently, (2) reduce information overload, and (3) better enable exploration of recommendations.
In the remainder of this section, we highlight the features of our design and the reasoning behind our design decisions.
%












\subsection{Utilize Screen Space More Efficiently}

\Intent{A primary goal is to utilize screen space, and we do it by overlaying the graph on the editor.}
A primary goal of Wandercode is to utilize screen space efficiently, since code editors already have many panes competing for a portion of the screen.
The graph of recommendations provided by Wandercode is overlaid on top of the code editor, as shown in \CalloutFigure{fig-wandercode}d.
The recommendations automatically update based on the text cursor's location in the code editor (i.e., which method of code is the text cursor in).
When a recommendation is clicked, that code location will open in a new tab document of the code editor. 

\Intent{The benefits of overlaying the recommendations are as follows.}
By providing the recommendations on top of the editor, the programmer does not need to manage any windows or panes within the code editor.
Since other tools place the recommendations inside a pane that is docked to the side or bottom of the editor, the programmer would have to decide what they want to view more: more of the editor, more of the recommendations, or more of some other view in their development environment.
For example, Reacher~\cite{LaToza2011VLHCC} and Prodet~\cite{Augustine2015ICSE} both take considerable screen space to display the graph of recommendations in a pane below the editor.
%
%
Given that code generally does not require much horizontal space~\cite{Short2019VLHCC}, Wandercode anchors the graph to the right edge of the code editor and constrains itself from ever taking up more than half the editor's width (one third if using a higher resolution screen).
In our informal testing of Wandercode, the graph of recommendations covered little to no code in the editor when using a 27 inch monitor with a standard screen resolution of 1920x1080 to view an open source codebase.
%

\subsection{Reduce Information Overload}

\Intent{Another goal is to prevent information overload, so we only display a small subset of the graph.}
Another goal of Wandercode is to reduce information overload, which is a paramount concern in code editors that already display many different types of information. 
To achieve this goal, Wandercode will initially display up to 6 recommendations on the screen at any given time.
The graph is organized in three portions: \CalloutFigure{fig-wandercode-graph}a shows methods that \emph{call} the method where the text cursor is placed, which is shown in \CalloutFigure{fig-wandercode-graph}b, and \CalloutFigure{fig-wandercode-graph}d shows methods that get \emph{called} by the current method.
%
%
Although there may be many methods, Wandercode only displays up to 3 methods on each side (i.e., 3 callers and 3 callees).
The recommendations are sorted based on their frequency, relative to the current code project, while discounting recommendations to other code projects (e.g., the standard library or an external library).
That is, if a code module is referenced more times in other code locations, then it will be ranked as more relevant.
While testing our recommender system, we observed that the recommendation rankings followed a long-tail distribution, with only 2--4 having a substantial rank.
Additionally, 6 recommendations will often fit in a user's working memory, the amount of information that a person can keep in their head at any given time, which has been estimated to be 3--9 items of information~\cite{Cowan2001, Miller1956}.

\Intent{Additionally, we only display minimal info about each recommendation.}
Another way that Wandercode strives to prevent overloading the programmer with information is by only displaying high-value information for each recommendation.
In particular, each recommendation shows the code location's method name and class name.
Other graph-based tools provide numerous types of additional information.
For example, Prodet displays the line number of the method call and an icon that represents what type of code module it is (e.g., method or class)~\cite{Augustine2015ICSE}.
%
%
We intentionally chose to not include this information in Wandercode's visualization since it is often irrelevant to tasks, may clutter the screen, and is easy for the programmer to retrieve by clicking the recommendation.
Furthermore, Wandercode's recommendations maintain a consistent appearance (e.g., color and size) since it does not encode any information in the appearance of recommendations.
In contrast, an existing tool, Coronado, increases the size of recommendations based on how relevant it is calculated to be~\cite{Ge2014VLHCC}, and Prodet colors the recommendations based on what class the recommendation belongs to~\cite{Augustine2015ICSE}.

\subsection{Better Enable Exploration of Recommendations}

\Intent{}
Wandercode also aims to enable programmers to explore all of the recommendations such that they can better comprehend the code and seek the locations most relevant to them.
To facilitate this, Wandercode provides a feature that will \emph{pin} the graph to the screen and prevents the recommendations from updating (\CalloutFigure{fig-wandercode-graph}c).
By using this feature, programmers can navigate to each caller or callee without the programmer losing their place in the graph. 
Once the programmer is finished exploring, the programmer can \emph{unpin} the graph, and it will continue updating when the text cursor is moved.
Other tools for recommending code either automatically update, thus preventing the programmer from investigating all of the recommendations, or the tool only updates recommendations when enacted (i.e., a button is pressed), which means the programmer will never be surprised by a new recommendation or that they might forget to use the tool entirely.

\Intent{If you need more, you can click the expand button.}
There are situations in which the programmer may want to view more recommendations.
To view more recommendations, the programmer can click the expand button depicted in \CalloutFigure{fig-wandercode-graph}c on the right, which will increase the number of recommendations up to 10.
\CalloutFigure{fig-wandercode-expanded} shows an already expanded graph.
%
%
Clicking the button again will hide the additional recommendations from the graph.

\DraftPageBreak{}


\section{Design Dimensions}
\label{sec-design}

\begin{table*}
\centering
\caption{%
Design dimensions that we identified during our design process of creating Wandercode. 
%
}
\vspace{1pt}
\includegraphics[scale=0.92]{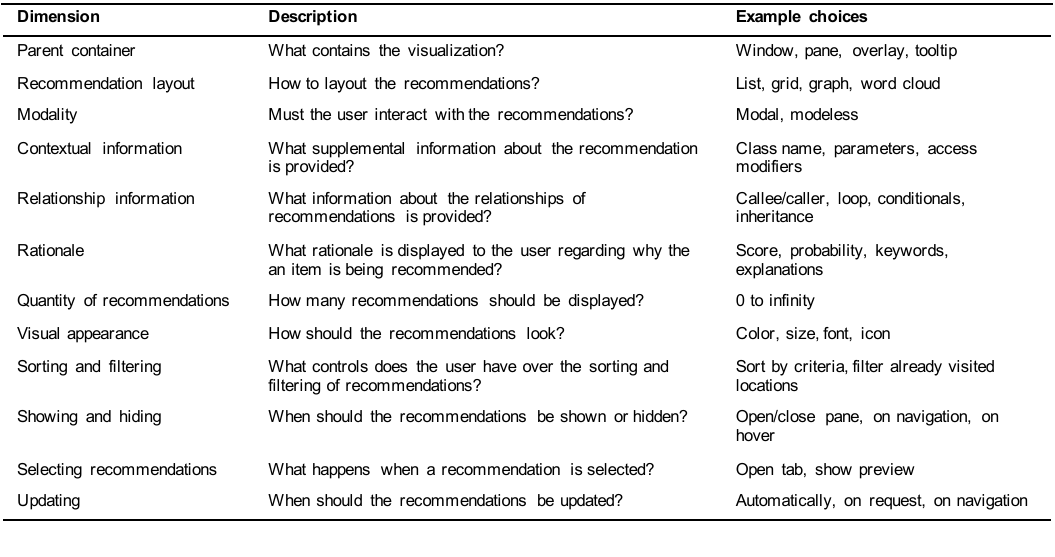}

\vspace{-16pt}
\label{tab-designdimensions}
\end{table*}

\Intent{}
During the design of Wandercode, we documented each design decision that we made, and developed a set of design dimensions.
%
These dimensions could aid another tool designer while designing a recommender system for code.
Shown in \CalloutTable{tab-designdimensions}, these dimensions cover the user interaction with the recommendations.
We exclude any details about the specific algorithm for generating the recommendations. 
%
%
We detail the benefits and consequences of the design decisions in the remainder of this section.

\subsection{Parent container and location}

\Intent{}
A particularly fundamental choice for a recommender system is what container to use and where to place it.
In a typical user interface, the options for container are a window, pane, overlay, or tooltip.
Placing the recommendations in their own window does separate it from the application and make it easy to fullscreen, but requires the user to then manage another window.
A pane is the standard approach that is supported in most code editors, balancing customization (moving and resizing) with window management activities.
Overlaying the recommendations over existing content reduces or eliminates window management but may hide essential information and get in the user's way.
A tooltip gives the user little to no control how and where the recommendations are displayed, but also uses little screen space except when the user wants to see it.
Another potential issue with tooltips is that the user may forget to take advantage of the feature since it is hidden by default.

\subsection{Recommendation layout}

\Intent{}
Choosing an efficient layout for the recommendations may have tremendous affect on how users interact with the information.
For example, they could be presented as a list, grid, word cloud, or multiple lists.
A list is simple, has been used by many tools, and doesn't require much screen space, but portrays less information.
Although a graph may portray more information, namely the structure of code, it could require more effort for the user to understand and it requires a lot of screen space.
A word cloud has also been studied but did not perform well in one evaluation~\cite{Ge2014VLHCC}.
Stacksplorer used multiple lists, portraying two layers of a graph, which may achieve a balance, but requires considerable screen space~\cite{Karrer2011UIST}.
%
%
A more radical option would be to present a 3D visualization with recommendations.
Although 3D visualizations have been investigated in other areas of software development (e.g., CodeCity~\cite{Wettel2007VISSOFT}), they have not yet been used for recommending code locations to programmers.

\subsection{Window Modality}

\Intent{}
The window modality describes how a user is able to interact with the recommendations.
Using a \emph{modal} container forces the user to interact with the recommendations window (e.g., like a popup dialog box would)
In contrast, a \emph{modeless} container allows the user to freely switch between the recommendations and other content on the screen.
Most recommendation systems use a modeless container, such that it does not disrupt the user's workflow and enables the user to switch between the code and the recommendations.
One likely benefit of modal is that a user can not accidentally ignore the recommendation.
It may also be easier to implement since the system does not need to handle all of the possible user interactions while the recommendation is displayed (e.g., what if the user deletes code that the tool was recommending).

\subsection{Contextual information}

\Intent{}
To help users evaluate and decide recommendations, tools often provide additional contextual information.
For code locations, this could be the class name, the parameters, the line number, the access modifier, who last modified the code, and the last time the user viewed the code location.
Additionally, a recommender system could display a snippet or preview of the code.
The challenging aspect is understanding which information would be beneficial to users, since displaying everything may lessen the utility of a recommendation.
Murphy-Hill and Murphy also argued the importance of providing contextual information to recommendations~\cite{Murphy-Hill2014RSSE}.

\subsection{Relationship information}

\Intent{}
Knowing how recommendations relate to one another is another way that tools have tried to make recommendations more useful.
This could be which methods call each other, often depicted by arrows connecting the recommendations, and loop or conditional info (e.g., this method is called in a loop from this method).
Even list tools can provide this information, such as Blaze~\cite{Kramer2012CHI} and Stacksplorer~\cite{Karrer2011UIST}.
Based on the results of our study, participants could have benefited from this information by the control tool separating the recommendations into two lists.
Given that studies have shown that programmers are more successful when they have structural information about the code~\cite{Ko2006TSE, LaToza2010ICSE}, this information could be valuable for a recommender system to provide.

\subsection{Rationale}

\Intent{}
Recommender systems may provide rationale to the user as to why the recommendation is being suggested.
For example, Suade displays a relevancy score from 0--1 and a reason about why that code location may be relevant (e.g., ``This method is called by \emph{GetCaretStatus()}'')~\cite{Warr2007ICSE}.
The PFIS recommender system displays a list of keywords for each recommendation regarding why the system believes that code location is relevant.
It may be difficult for users to make sense of the rationale though; for example, a user may not know how to interpret the difference between a a relevancy score of 0.65 and 0.9.
An alternative is to provide a textual explanation as to why the recommendation is relevant.

\subsection{Quantity of recommendations}

\Intent{}
Another design dimension is the question of how many recommendations to provide to a user.
Considering that software projects may have hundreds of thousands of code locations, it is not beneficial to display all of them.
Additionally, the quantity of recommendations must balance the need for screen space and effort for the user to understand the recommendations.
For Wandercode, we limited the number of recommendations to 6, but provided a button that will expand the number to 10.
Our reasoning took into account screen space, human memory capacity~\cite{Cowan2001, Miller1956}, and the long-tail distribution that we observed from our recommendation system.

\subsection{Visual appearance}

\Intent{}
The visual appearance of recommendations may also impact how a user interacts with the recommendations.
The recommendation's color, size, font, and icons are all elements that can be designed to further help the user.
Additionally, recommendation systems can encode contextual information in the appearance as well.
For example, Prodet colors recommendations based on class~\cite{Augustine2015ICSE} and Coronado sizes recommendations based on their relevancy~\cite{Ge2014VLHCC}.

\subsection{Sorting \& Filtering}

\Intent{}
Although the backend of the recommendation system determines which recommendations to display, the visualization can have some control over sorting and filtering as well.
For example, the tool could provide buttons for sorting or filtering the recommendations based on different criteria.
Codebroker allows the user to hide classes that are deemed not relevant by the user~\cite{Ye2002ICSE}.
Mylar hides any code location not recently visited by the user and allows the user to collapse a folder or file to hide those recommendations~\cite{Kersten2006FSE}.

\subsection{Showing and hiding}

\Intent{}
When to show and hide the recommendations to the user is a fundamental aspect of the interaction design, however, most tools have not investigated the most efficient means of doing so.
Possible options for this design dimension include opening/closing a pane, showing based on an event (e.g., a navigation or a pause in activity), or showing when hovering the mouse over a code element (like a tooltip).
Every existing tool that we investigated shows the recommendations continuously in a pane and allows the user to hide them by closing the pane.

\subsection{Updating}

\Intent{}
The choice of when to update recommendations is another design decision that may drastically change how users will interact with the tool.
A particularly common approach is for recommendation systems to continuously update the recommendations (e.g., Mylar~\cite{Kersten2006FSE} and the PFIS recommender~\cite{Piorkowski2012CHI}).
However, this approach does not enable the user to try out each recommendation without the risk that they may lose some recommendations.
Another approach is to only update the recommendations when the user requests it, often by clicking a button to get recommendations (e.g., Suade~\cite{Warr2007ICSE}).
For Wandercode, we chose a hybrid approach by having the default behavior be that the recommendations continuously update, but also provide a button that pins the current recommendations on screen and prevents them from updating.

\subsection{Selecting a recommendation}

\Intent{}
The main benefit of providing recommendations is to allow the user to do something with it, namely, navigate to that location.
What occurs when the user clicks on the recommendation must be decided.
Common options include opening a new tab document or navigating the current tab document to the code location.
Other possible options include showing a popup window of the code, like a preview, or even splitting the current tab document in two to show a comparison of the code (e.g., Visual Studio's Peek Definition feature).

\DraftPageBreak{}


\section{Evaluation Method}
\label{sec-method}

\Intent{Intro to method.}
To address our research questions (see Introduction), we conducted a laboratory study of programmers engaged in comprehension tasks.
The study had two treatments: the \emph{control} treatment which used a list-based recommender system based on traditional approaches (described below), and the \emph{Wandercode} treatment which used the Wandercode graph-based recommender system.
Both tools were implemented as extensions to the Atom code editor, and all other features were consistent for both treatments.
The study had a within-subjects design in which each participant experienced both treatments.
To account for order effects, we blocked and balanced based on the treatment orderings.

\subsection{Participants}
\label{sec-participants}

\Intent{Here is what our participants were like.}
Our participants consisted of 10 Java programmers (7 male, 3 female), made up of 8 graduate students, 1 undergraduate student, and 1 professional programmer. 
%
%
They reported, on average, 5.3 years of programming experience ($\mathit{SD} = 2.45$) and 2.3 years of Java programming experience ($\mathit{SD} = 0.67$).
All participants reported having experience with modern code editors, such as Eclipse and Sublime.

\subsection{Control Treatment}

\begin{figure}
\centering
\includegraphics[scale=0.92]{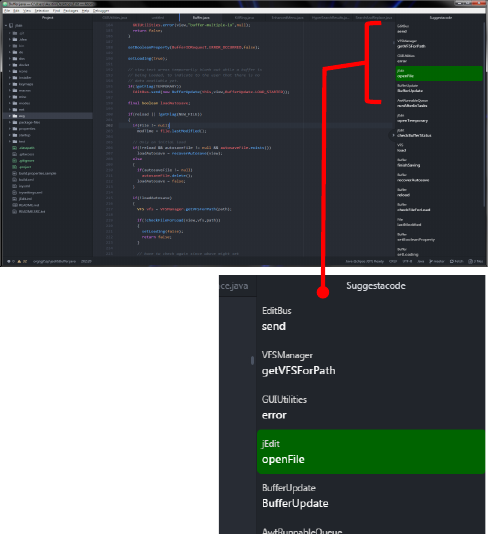}

\vspace{-2pt}
\caption{%
Atom with the control extension displaying a list of recommendations.
The list is sorted by top to bottom by relevancy, and each recommendation displays the method name and the class name.
%
The highlighted recommendation is where the mouse is currently hovering.
%
}
\label{fig-suggestacode}
\end{figure}

\Intent{The control treatment was like this.}
To compare Wandercode to traditional recommender systems that utilize a list-based visualization, we designed and implemented such a tool, as seen in \CalloutFigure{fig-suggestacode}.
It was presented to participants in the same way as Wandercode, a tool for providing recommendations, and it was called \emph{Suggestacode}.
We drew inspiration from other tools, such as Navtracks~\cite{Singer2005ICSM}, Suade~\cite{Warr2007ICSE}, the PFIS recommender~\cite{Piorkowski2012CHI}, and Mylar~\cite{Kersten2006FSE}, which provide a list of recommendations of which code modules may be of interest to the programmer.
%
%
%
The control uses the same recommender system as Wandercode and differs only in how the recommendations are displayed.
Instead of displaying the recommendations in a graph, the recommendations are displayed in a vertical list that is docked to the right side of the code editor.
Each recommendation contains the name of the code method and the name of the class that the method is contained in.
The recommendations automatically update when the programmer moves the text cursor to another method of code.
When a recommendation is clicked, it will open that code location in the editor.

%
%
%
%


\subsection{Comprehension Tasks}
\label{sec-tasks}

\Intent{Here are their tasks.}
For our study, the participants answered comprehension questions regarding the code structure of a large open source Java project, jEdit~\footnote{http://www.jedit.org/}.
Such comprehension is fundamental to virtually any software development task, which requires the programmer to understand the relationships between code modules to be successful in modifying code~\cite{Ko2006TSE, LaToza2010ICSE}.
Our study design took inspiration from two prior studies on the effects of recommender systems for code locations~\cite{Karrer2011UIST, LaToza2011VLHCC}, in particular, by having tasks that involve answering questions regarding the code's structure.
Each participant performed eight tasks involving answering a specific question about the relationships between code methods and classes.
The question was provided to them on paper along with a starting point in the jEdit project and rationale as to why they might need to know this information.
An example of a question is, ``Imagine you need to modify the parameters of the method \Code{setWholeWord()}. However, doing so would cause errors in any method that calls \Code{setWholeWord()}. Which methods would need to be fixed? (There are two.)''
They were allowed to use any navigation feature available to them in Atom.
If they did not complete the task within 10 minutes, we asked them to continue to the next task (this only occurred once).
The tasks were divided into two sequences of four (one sequence for each treatment).
For each four-task sequence, there were two \emph{upstream} questions and two \emph{downstream} questions, referring to the direction in which the answer was in the code's call graph.


\subsection{Procedure}
\label{sec-procedure}

\Intent{Here is exactly what they did for an hour.}
Each participant took part in an individual session that lasted approximately 1 hour.
All participants began the session by filling out a background questionnaire and receiving an introduction to the Atom code editor and jEdit codebase.
We demonstrated how Atom's code navigation features work, such as searching within a file, searching within a project, and finding all references of a code element. 
%
%
Five randomly selected participants used the control extension first, and the other five used the Wandercode extension first.
Each 4-task sequence began with an introduction to the extension that the participant would be using for that sequence.
Next, the participant performed the 4-task sequence, completing each task before beginning the next.
We asked each participant to ``think aloud'' as they worked.
Once the participant had finished each task, they completed a cognitive-load questionnaire (details below).
At the end of each 4-task sequence, the participant completed a usability questionnaire regarding the extension they used.
After both 4-task sequences were completed, all participants took part in a semi-structured interview in which they discussed their experiences using the two extensions.


\subsection{Data Collection}
\label{sec-data-collection}

\Intent{Here's the data we collected.}
The data collected for the study comprised screen-capture video and audio of the participants as well as their questionnaire responses.
For the cognitive load questionnaire, we used a well-validated instrument based on Cognitive Load Theory~\cite{Paas2003EduPsych}.
The instrument features a 7-point Likert question that measures a person's overall cognitive load during a task.
Prior work showed that the instrument does not interfere with task performance, is sensitive to small differences in workload, and is reliable~\cite{Paas1994PMS}.
Moreover, the instrument has been shown to highly correlate with more-complex self-reporting instruments (e.g., NASA TLX)~\cite{Windell2007AERA} as well as with physiological sensors (e.g., heart rate)~\cite{Paas1994PMS}.

\DraftPageBreak{}


\section{Results}
\label{sec-results}



\begin{figure}
\centering
\includegraphics[scale=0.9]{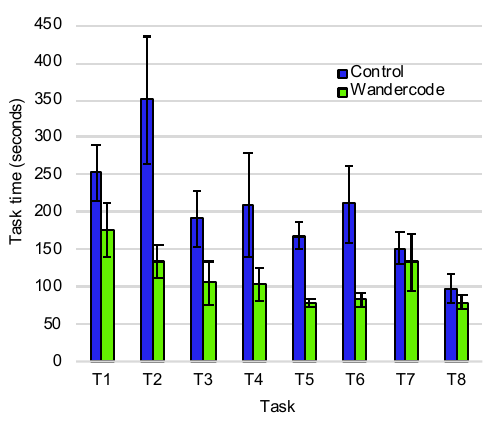}

\vspace{-5pt}
\caption{%
Wandercode users completed tasks significantly faster than control users (smaller bars are better).
Whiskers denote standard error.
%
}
\label{fig-wandercode-results-time}
\end{figure}

\begin{figure}
\centering
\includegraphics[scale=0.9]{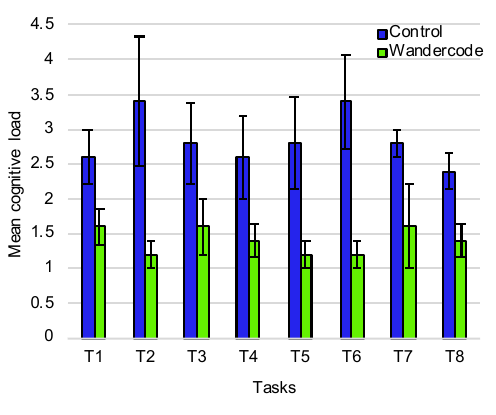}

\vspace{-10pt}
\caption{%
Wandercode users reported significantly lower cognitive load than control users (smaller bars are better).
Whiskers denote standard error.
%
}
\label{fig-wandercode-results-cognitiveload}
\end{figure}

\begin{figure}
\centering
\includegraphics[scale=0.9]{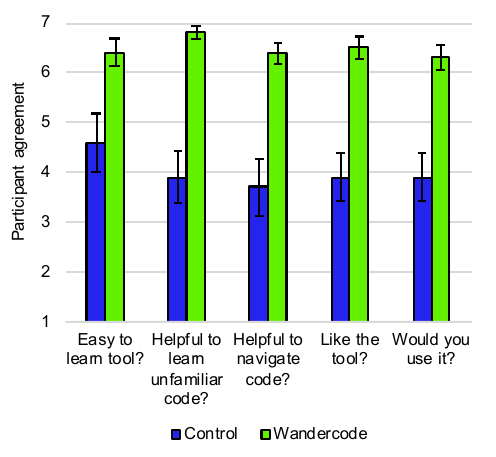}

\vspace{-8pt}
\caption{%
Participants reported significantly higher opinions of Wandercode than of the control tool (larger bars are better).
Whiskers denote standard error.
%
}
\label{fig-wandercode-results-usability}
\end{figure}

\subsection{RQ1 Results: Time on Task}


\Intent{Wandercode significantly reduced the time on task.}
As \CalloutFigure{fig-wandercode-results-time} shows, participants using the Wandercode extension completed tasks considerably faster bugs than those using the control extension.
In fact, Wandercode users completed every task faster, on average.
Indeed, the results of a Mann--Whitney $U$ test revealed that Wandercode users completed tasks significantly faster than control users ($U = 7$, $Z = 2.57$, $p < 0.05$).


\subsection{RQ2 Results: Cognitive Load}


\Intent{Wandercode had significantly lower cognitive load.}
As \CalloutFigure{fig-wandercode-results-cognitiveload} shows, Wandercode users reported considerably lower cognitive loads than control users.
The Wandercode users reported a lower cognitive load than control users for every task, on average.
For the task that the most participants answered incorrectly, Task 5, control users reported a cognitive load that was nearly 3 times higher than Wandercode users, on average. 
The results of a Mann--Whitney $U$ test showed that Wandercode users reported a significantly lower cognitive load than control users ($U = 0$, $Z = 3.3$, $p < 0.01$).


\subsection{RQ3 Results: Opinions of the Participants}

\Intent{To assess participant opinions of the tools, we asked them these questions.}
To assess participants' opinions of the two tools, each participant responded to the following questions at the end of the task sequences (7-point Likert scale):
\begin{itemize}
\setlength\itemsep{-0.1em}

\item
How difficult/easy was this tool to learn?

\item
How unhelpful/helpful do you think this tool is for learning about unfamiliar code?

\item
How unhelpful/helpful do you think this tool is for navigating code?

\item
How much did you dislike/like using this tool?

\item
If this tool was made available to you, how often would you use it for your coding tasks?

\end{itemize}

\Intent{Wandercode had significantly better opinions.}
As \CalloutFigure{fig-wandercode-results-usability} shows, users reported significantly higher opinions of Wandercode than the control tool.
In fact, every participant reported a higher or equivalent score for Wandercode than they did for the control.
%
%
The results of a Mann--Whitney $U$ test showed that participants reported significantly higher opinions of Wandercode than the control tool ($U = 0$, $Z = 3.74$, $p < 0.001$).
Additionally, when asked which tool they preferred, 9 of the 10 participants said Wandercode, and one participant said ``it depends'' on the situation.

\DraftPageBreak{}


\section{Discussion}
\label{sec-discussion}

\Intent{}
Overall, the results of our empirical evaluation of Wandercode were favorable. 
Participants using Wandercode in far less time than when using the control tool (RQ1 results).
Additionally, participants using Wandercode were able to complete the tasks with far less effort, as their self-reported cognitive load shows (RQ2 results).
Moreover, participants rated Wandercode to be far easier to use and learn, and all but one of the participants outright preferred Wandercode over the control tool (RQ3 results).
%
%
In the remainder of this section, we describe qualitative observations of the participants using the tools and the limitations of our study. 

\subsection{Qualitative Observations}

\subsubsection{Wandercode displayed structure}

\Intent{}
It was evident from observing the participants that Wandercode's graph was providing structural information that can be tedious to obtain otherwise.
P6 was especially fast with using Wandercode, taking less than 90 seconds on several of his tasks.
In particular, P6 was able to complete Task 3 in just 59 seconds with minimal navigations with Wandercode.
He began the task with a textual search to find a starting point.
Once Wandercode's graph was displayed, he traced the graph with the mouse and proclaimed, "done."
Later during his interview session, P6 confessed that he really liked Wandercode thanks to it telling him the callers and callees of a method.

\Intent{}
Other participants also gave their thoughts on why they liked Wandercode.
P7 summarized his opinion on the benefits of Wandercode:

\begin{ParticipantQuote}

\item[P7:]
``When I am navigating new code, what I need is a picture in my mind of the relationships between classes and methods, and who is calling what, and it takes a reasonable amount of time to figure that out. This graph already creates that image.''

\end{ParticipantQuote}

\noindent
Several participants also commented about the visualization being beneficial:

\begin{ParticipantQuote}

\item[P3:]
``It is very obvious to understand.''

\item[P4:]
``The graph gave a better picture.''

\item[P5:]
``It is easy to see and visualize... I can easily see the functions.''

\end{ParticipantQuote}

\noindent
Other participants described how the structural relationships helped them be more efficient:

\begin{ParticipantQuote}

\item[P9:]
``With the graph, I didn't need to look into every method and look at the code.''

\item[P10:]
``The graph gives a clear distinction which functions are called and which functions it's calling''

\end{ParticipantQuote}
Another positive sentiment was from P3 after she learned how to use Wandercode, she said "ah, now these [tasks] will be much easier."


\subsubsection{The list could be useful too, at times}

\Intent{The list helped but was not efficient. Here is P4s episode.}
While using the control, participants did make use of the recommendations, but were not nearly as efficient in doing so.
For example, P4 struggled while using the control tool for Task 6.
%
%
After several attempts to skim the code and perform text searches, he quickly scrolled through the list of recommendations and hovered on the second recommendation, \Code{loadMenuItem}, for several seconds.
%
Eventually he moved his attention back to the open file of code and tried another text search.
Getting frustrated, P4 said, ``there is no method createMenuItems!''
He then went back to looking at the recommendations and quickly noticed \Code{createMenuItems} was the 5th item in the list and excitedly exclaimed, "ooh, ooh!"
Having found the other relevant method, he navigated back and forth between the two methods to understand how they interact.
Finally, after 4 and a half minutes, he completed the task, despite spending less than 90 seconds on each of the earlier tasks with Wandercode.
P4 later expressed his dissatisfaction with the list:

\begin{ParticipantQuote}

\item[P4:]
``I do not know what to do with this... I'm just going to use Ctrl+F.''

\end{ParticipantQuote}

\Intent{Here are the complaints people had.}
Other participants expressed their frustrations about the list of recommendations as well.
In particular, the participants were unclear about why the recommendations were being chosen:

\begin{ParticipantQuote}

\item[P3:]
``I do not know what kind of information it will give.''

\item[P1:]
``The list was a bit opaque, it was just listing things. it seemed arbitrary.''

\item[P5:]
``When I see the list, I have no idea where the list comes from.''

\end{ParticipantQuote}

\Intent{Here are some positive things.}
In stark contrast, some participants expressed positive sentiments after using the list:

\begin{ParticipantQuote}

\item[P2:]
``Usually what I do is start by searching. Instead of that, with this [list], whenever I open up a file I think this would be more helpful.''


\item[P9:]
``It would help, but not like the graph.''

\item[P10:]
``It will reduce the time it takes to do manually.''

\end{ParticipantQuote}

\noindent
%
Participants also provided suggestions on how the list could be improved.
6 of the 10 participants suggested either splitting the list into two lists, one for callees and one for callers, or to draw arrows between the recommendations to show their relationship.
P8 stated that he did not want it to take up so much space and suggested making it show in a tooltip when hovering over a method.

\subsubsection{Improvements to Wandercode}

Although participants were highly favorable of Wandercode, they did suggest ways it could be improved.
P4 and P6 asked for features to view more methods or even to see the entire graph in a fullscreen view.
Four participants suggested ways to utilize screen space even more efficiently.
P1 asked to make the graph vertical while P7 and P8 asked for a feature to show the graph temporarily, such as a keyboard shortcut.
P9 wanted to move the graph around the screen to better position it.
Finally, P4 also suggested needing a way to know if methods were being filtered or if all of them were displayed.

\subsection{Limitations}

\Intent{List limitations here.}
Our user study has several limitations inherent to laboratory studies of programmers.
First, our sample of programmers, consisting of mostly graduate students, may not be representative of all programmers.
Second, the tasks are questions that may not be ecologically valid in a broader software development context; however, we based them on prior studies of relevant tools that used empirical evidence for the basis of the questions.
Third, reactivity effects (e.g., participants consciously or unconsciously trying to please the researchers) may have occurred; however, we tried to minimize the effects by presenting the two tools as extensions to Atom, and did not disclose that they were our invention.
Finally, order effects of the tool and tasks may have affected our observations, but we counterbalanced the tool order to control for the effect with respect to our statistical tests and kept the task order the same for all participants.

\DraftPageBreak{}


\section{Conclusion}
\label{sec-conclusion}

\Intent{In this paper, we presented Wandercode.}
In this paper, we presented the novel Wandercode interaction design for recommender systems that aid programmers in understanding and navigating large codebases.
In particular, Wandercode aims to help programmers by reducing information overload, easing exploration, and conserving screen space.
An evaluation study comparing our Wandercode prototype to a representative list-based design made the following key findings:

\begin{itemize}
\setlength\itemsep{-0.1em}
\item RQ1 (task time): Programmers using Wandercode completed comprehension tasks significantly faster than those using the control tool.

\item RQ2 (cognitive load): Programmers using Wandercode reported significantly lower cognitive load during comprehension tasks than those using the control tool.

\item RQ3 (user opinions): Programmers rated Wandercode to be significantly easier to use and learn than the control tool.

\end{itemize}

\Intent{Future work and sunset.}
We hope that Wandercode marks a substantial step toward supporting the arduous task of comprehending and navigating massive codebases.
Our study findings suggest future work to further support exploring the structure of code by enabling the programmer to view a full-screen graph to get a ``big picture'' of the code.
In this view, the programmer could keep track of interesting recommendations by starring them, thus enabling the programmer to investigate in more detail at another time.
Furthermore, we will run a field study of Wandercode to understand how programmers utilize the recommendations in a more realistic setting by releasing Wandercode to the public and collecting usage data.
These ideas have considerable potential to dramatically improve the productivity of programmers and reduce their mental burden when faced with navigating code.



\balance{}

\bibliographystyle{SIGCHI-Reference-Format}
\bibliography{refs}

\end{document}